\documentclass[sigconf]{acmart}
\AtBeginDocument{%
  }
\setcopyright{acmlicensed}
\copyrightyear{2024}
\acmYear{2024}
\setcopyright{rightsretained}
\acmConference[VRST '24]{30th ACM Symposium on Virtual Reality Software and Technology}{October 9--11, 2024}{Trier, Germany}
\acmBooktitle{30th ACM Symposium on Virtual Reality Software and Technology (VRST '24), October 9--11, 2024, Trier, Germany}\acmDOI{10.1145/3641825.3689526}
\acmISBN{979-8-4007-0535-9/24/10}

\begin{document}

\title{Digital Eyes: Social Implications of XR EyeSight}


\author{Maurizio Vergari}
\affiliation{%
  \institution{Quality and Usability Lab, TU Berlin}
  \city{Berlin}
  \country{Germany}
}
\email{maurizio.vergari@tu-berlin.de}

\author{Tanja Kojić}
\affiliation{%
 \institution{Quality and Usability Lab, TU Berlin}
\city{Berlin}
\country{Germany}}
\email{tanja.kojich@tu-berlin.de}

\author{Wafaa Wardah}
\affiliation{%
  \institution{Quality and Usability Lab, TU Berlin}
  \city{Berlin}
  \country{Germany}
}
\email{wafaa.wardah@tu-berlin.de}

 \author{Maximilian Warsinke}
\affiliation{%
  \institution{Quality and Usability Lab, TU Berlin}
  \city{Berlin}
  \country{Germany}}
\email{warsinke@tu-berlin.de}

\author{Sebastian M\"oller}
\affiliation{%
  \institution{Quality and Usability Lab, TU Berlin}
  \city{Berlin}
  \country{Germany}
}
\email{sebastian.moeller@tu-berlin.de}

\author{\hbox{Jan-Niklas Voigt-Antons}}
\affiliation{%
  \institution{Immersive Reality Lab, \hbox{Hochschule Hamm-Lippstadt}}
  \city{Lippstadt}
  \country{Germany}
}
\email{Jan-Niklas.Voigt-Antons@hshl.de}

\author{Robert P. Spang}

\affiliation{%
\institution{Quality and Usability Lab, TU Berlin}
\city{Berlin}
\country{Germany}
}
\email{spang@tu-berlin.de}

\renewcommand{\shortauthors}{Vergari et al.}

\begin{abstract}
  The EyeSight feature, introduced with the new Apple Vision Pro XR headset, promises to revolutionize user interaction by simulating real human eye expressions on a digital display. This feature could enhance XR devices' social acceptability and social presence when communicating with others outside the XR experience. In this pilot study, we explore the implications of the EyeSight feature by examining social acceptability, social presence, emotional responses, and technology acceptance. Eight participants engaged in conversational tasks in three conditions to contrast experiencing the Apple Vision Pro with EyeSight, the Meta Quest 3 as a reference XR headset, and a face-to-face setting. Our preliminary findings indicate that while the EyeSight feature improves perceptions of social presence and acceptability compared to the reference headsets, it does not match the social connectivity of direct human interactions. 
\end{abstract}

\begin{CCSXML}
<ccs2012>
   <concept>
       <concept_id>10003120.10003138</concept_id>
       <concept_desc>Human-centered computing~Ubiquitous and mobile computing</concept_desc>
       <concept_significance>500</concept_significance>
       </concept>
   <concept>
       <concept_id>10010147.10010371.10010387.10010393</concept_id>
       <concept_desc>Computing methodologies~Perception</concept_desc>
       <concept_significance>500</concept_significance>
       </concept>
 </ccs2012>
\end{CCSXML}

\ccsdesc[500]{Human-centered computing~Ubiquitous and mobile computing}
\ccsdesc[500]{Computing methodologies~Perception}

\keywords{Extended Reality, Social Acceptability, Social Presence}
  


\maketitle

% %%%%%%%%%%%%%%%%%%%%%%%%%%%%%%%%%%%%%%%%%%%%%%%%%%%%%%%%%%%%%%%%%%%%%%%%%%%

\section{Introduction and Related Work}
% %%%%%%%%%%%%%%%%%%%%%%%%%%%%%%%%%%%%%%%%%%%%%%%%%%%%%%%%%%%%%%%%%%%%%%%%%%%
We are witnessing extraordinary improvements in the Extended Reality (XR) field. 
Most recently, one device that is capturing a lot of attention is the Apple Vision Pro (AVP), a cutting-edge XR headset that includes, among all, an innovative feature called "EyeSight". This feature aims to model the eyes of the users and displays them on a front display to others. Eyes are known to be able to contribute to the uncanny valley effect in other fields. Moreover, this kind of visual feedback given to the observers could affect the social acceptability of XR headsets and increase social presence. Social Acceptability can be defined as "the phenomenon of judging a technology introduced in the future from the comfort or discomfort of both the performer and the observers" \cite{distler2018acceptability}. At the same time, Social Presence is the "psychological state in which the individual perceives themselves as existing within an interpersonal environment"\cite{blascovich2002social}. Studies on Social Acceptability from the observer's perspective argue that it depends on the situation, the environment, and the visibility of the interaction \cite{ alallah2018performer, eghbali2018social}. Research on Social Presence explores how individuals perceive and engage with different technologies. With this preliminary study, we aim to make an initial assessment of the social implications of the EyeSight feature. The following research questions arise: 1. How does EyeSight influence observers' Social Acceptability of XR? 2. How does EyeSight influence observers' Social Presence? 3. What factors contribute to the observers' impression of EyeSight?

% %%%%%%%%%%%%%%%%%%%%%%%%%%%%%%%%%%%%%%%%%%%%%%%%%%%%%%%%%%%%%%%%%%%%%%%%%%%
\section{Method}
% %%%%%%%%%%%%%%%%%%%%%%%%%%%%%%%%%%%%%%%%%%%%%%%%%%%%%%%%%%%%%%%%%%%%%%%%%%%

        Eight participants, five men and three women, took part in this pilot study. The average age was 28.25 (SD=2.25, min=26, max=31). The user study for this research was done in a lab setting, where the participants were invited, each at a different time slot. To create a conversational setting for each trial, participants engaged in survival tasks (\cite{rec1301p}). They were presented with five different items and had to reach a consensus on essential items for survival in a hypothetical scenario. The experimenter also had five items, and together, they needed to discuss which four of the ten items would be most helpful for their imaginary situation. This task was designed to elicit a lively and realistic conversation. The average conversation took about three minutes per trial (as suggested in \cite{rec805p}). Participants were given 90 seconds to memorize the items on their list before starting the discussion. Participants interacted freely with the experimenter who was wearing a headset or not, depending on the condition. The moderator was the one who always wore the AVP or the reference headset -Meta Quest 3 (MQ3) for the applicable conditions. There were three experimental conditions: Face-to-face, MQ3-to-face, and APV-to-face conversation. The conditions were randomized, and the tasks were always run from tasks 1 to 3. After each condition, the participants were asked to answer web-based questionnaires encompassing the SAM \cite{bradley1994measuring}, Social Acceptability from observer's perspective \cite{eghbali2018social}, and NMSPI questionnaires \cite{NMPI6743}. After all the conditions had been played and rated, the participants were asked to rate the AVP EyeSight feature with the Technology Acceptance Model (TAM) questionnaire \cite{davis1989technology} and to provide feedback about how comfortable this feature made them feel. Given the small sample size of this pilot, inferential statistics are not applied and these initial results are discussed on a descriptive basis only.

% %%%%%%%%%%%%%%%%%%%%%%%%%%%%%%%%%%%%%%%%%%%%%%%%%%%%%%%%%%%%%%%%%%%%%%%%%%%

\setlength{\belowcaptionskip}{-6mm}

      \begin{figure}[tbph]
            \centering
            \includegraphics[width=\linewidth]{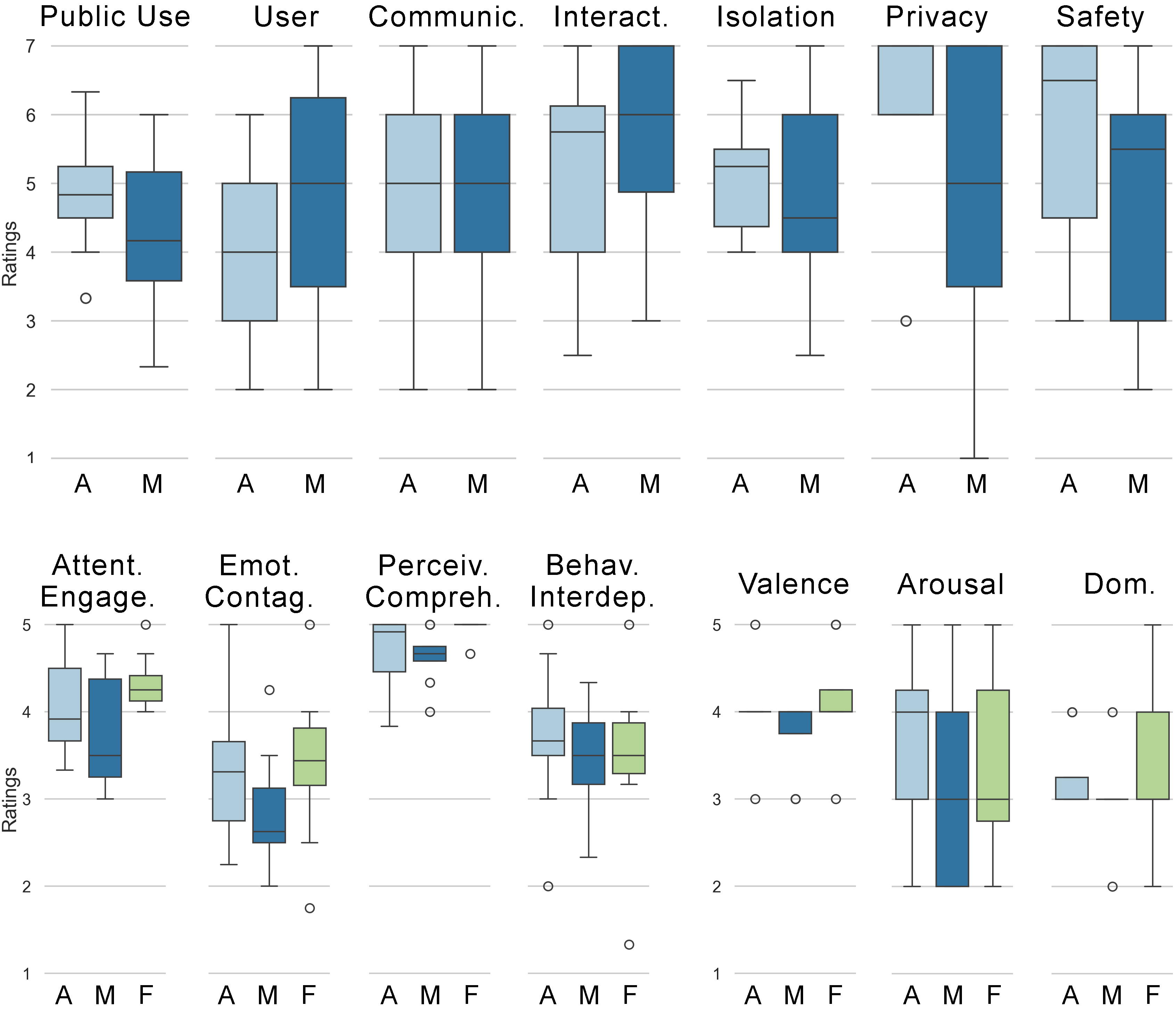}
            \caption{Contrasts between the Apple Vision Pro with EyeSight (A), the Meta Quest 3 (M), and a direct face-to-face (F) conversational setting. Top row: Seven Social Acceptability contrasts. Bottom row from left to right: Social Presence sub-dimensions, and Valence, Arousal, and Dominance.}
            \label{fig:all_per_trial_combined}
        \end{figure}

% %%%%%%%%%%%%%%%%%%%%%%%%%%%%%%%%%%%%%%%%%%%%%%%%%%%%%%%%%%%%%%%%%%%%%%%%%%%

\section{Findings and Conclusion}
% %%%%%%%%%%%%%%%%%%%%%%%%%%%%%%%%%%%%%%%%%%%%%%%%%%%%%%%%%%%%%%%%%%%%%%%%%%%

    % social acceptability
    
    Given the pilot nature of this work, results are initial findings and generalizability needs to be confirmed in future studies. However, the preliminary findings on social acceptability (see Fig. \ref{fig:all_per_trial_combined}-top) suggest that the EyeSight feature may have potential in bridging the gap between digital and physical interactions. It tends to be perceived slightly more favorable for public settings compared to the MQ3, possibly due to its ability to simulate eye contact. This feature tends to be perceived slightly more favorably for public settings compared to the MQ3, possibly due to its ability to simulate eye contact, which could potentially reduce the perceived barrier between the XR user and the observer, thereby enhancing the social acceptability of XR technologies. However, the AVP, despite scoring higher in aspects like privacy concerns, does not seem to improve the naturalness of interactions as compared to traditional headsets. This highlights the challenge XR devices face in mimicking real human interactions. The variance in responses concerning the isolation of the user and the naturalness of body movements hints that individual differences in personal comfort with XR technology might play a significant role. These differences highlight the necessity for further studies involving a more extensive and diverse participant base to better generalize these initial findings across different populations. The analysis of Social Presence (see Fig. \ref{fig:all_per_trial_combined}-bottom) yielded intriguing initial insights, notably that interactions involving the AVP are perceived as more engaging and comprehensible than those with the MQ3. This perceived improvement could be attributed to the EyeSight feature's enhancement of the psychological impact of being present with another person, even when separated by digital interfaces. Despite these preliminary advancements, the direct face-to-face setting still appears superior in fostering a robust sense of social presence. These findings suggest that while XR technologies have made significant strides in enhancing remote or augmented communication, it seems that remains an irreplaceable quality to physical presence. The emotional responses measured through the SAM model (see Fig. \ref{fig:all_per_trial_combined}-bottom), indicate minimal variation in valence and arousal across different settings, which could suggest a baseline emotional neutrality towards the technology. However, the dominance dimension reveals a preference for face-to-face interaction, which could be interpreted as participants feeling more in control or at ease in non-mediated interactions. Participants generally viewed the EyeSight feature favorably, indicating a high perceived usefulness and a positive attitude towards the technology. This preliminary acceptance is promising for the potential integration of such technologies in everyday scenarios. However, the discomfort ratings and qualitative feedback reveal a mixed reception, with some participants noting the unnatural appearance and behavior of the digital eyes.

% %%%%%%%%%%%%%%%%%%%%%%%%%%%%%%%%%%%%%%%%%%%%%%%%%%%%%%%%%%%%%%%%%%%%%%%%%%%

\bibliographystyle{ACM-Reference-Format}
\bibliography{references}

\end{document}